\documentclass[final,5p,times,twocolumn]{elsarticle}

\usepackage{graphics}
\usepackage{xcolor}
\usepackage{amsmath}
\usepackage{mathtools, cuted}
\usepackage{lipsum, color}
\usepackage{amssymb}
\usepackage{braket}
\usepackage{multicol}
\usepackage{blindtext}
\usepackage{float}
\usepackage[flushleft]{threeparttable}
\usepackage{tikz}
\newcommand*\circled[1]{\tikz[baseline=(char.base)]{
            \node[shape=circle,draw,inner sep=0.5pt] (char) {#1};}}
\usepackage{lineno}
\journal{Journal of Applied Crystallography}

\begin{document}

\begin{frontmatter}

\title{The Linear Phase Correction of MIEZE with Magnetic Wollaston Prisms}
\tnotetext[label0]{This manuscript has been authored by UT-Battelle, LLC, under contract DE-AC05-00OR22725 with the US Department of Energy (DOE). The US government retains and the publisher, by accepting the article for publication, acknowledges that the US government retains a nonexclusive, paid-up, irrevocable, worldwide license to publish or reproduce the published form of this manuscript, or allow others to do so, for US government purposes. DOE will provide public access to these results of federally sponsored research in accordance with the DOE Public Access Plan (http://energy.gov/downloads/doe-public-access-plan).}

\author[label1]{Fankang Li\corref{cor1}}
\address[label1]{Neutron Technologies Division, Oak Ridge National Laboratory, Oak Ridge, TN, USA, 37831}
\cortext[cor1]{Corresponding author}
\ead{frankli@ornl.gov}

\begin{abstract}
I propose the use of two magnetic Wollaston prisms to correct the linear Larmor phase aberration of MIEZE, introduced by the transverse size of the sample. With this approach, the resolution function of MIEZE can be optimized for any scattering angle of interest. The optimum magnetic fields required for the magnetic Wollaston prisms depend only on the scattering angle and the frequency of the RF flippers and they are independent of the neutron wavelength and beam divergence, which makes it suitable for both pulsed and constant wavelength neutron sources.
\end{abstract}

\begin{keyword}
%% keywords here, in the form: keyword \sep keyword
MIEZE, linear phase correction, wide angle spin echo, intensity modulations, time and space focusing
%% MSC codes here, in the form: \MSC code \sep code
%% or \MSC[2008] code \sep code (2000 is the default)
\end{keyword}

\end{frontmatter}

%%
%% Start line numbering here if you want
%%
%%\linenumbers
%% main text

\section{MIEZE and its Larmor Phase Aberration}

The dynamical properties of materials are often the key to understanding their macroscopic properties. To investigate the slow dynamics, a technique that can reach the time scales in the range of a few tens of ns to 1 $\mu s$ with a matching length scale in the range of a few $nm$ to several hundred $nm$ is greatly desired. The Neutron Spin Echo (NSE) technique satisfies this demand by employing the Larmor labeling of neutron spins, which decouples the energy resolution of a neutron instrument from the flux and thus provides us with another approach to achieve ultra-high-resolution \cite{RN2360}. To implement NSE, two static magnetic fields are required with one on each side of the sample. The neutron spin will precess continuously inside the two magnets through the sample region, where the neutron spin is flipped and the Larmor phase accumulation will be reversed. But for certain situation, such as the samples that would depolarize the neutron or require high magnetic field, it is very challenging for NSE because it is difficult for neutrons to maintain the polarization through the sample. Even though paramagnetic \cite{RN2363}, ferromagnetic \cite{RN2362} and intensity modulated NSE \cite{RN2361} have been developed, either they cannot be applied for depolarizing sample environment or they are too complicated to be routinely operated.

\begin{figure*}[ht]
	\centering
	\includegraphics[width=0.8\linewidth, clip]{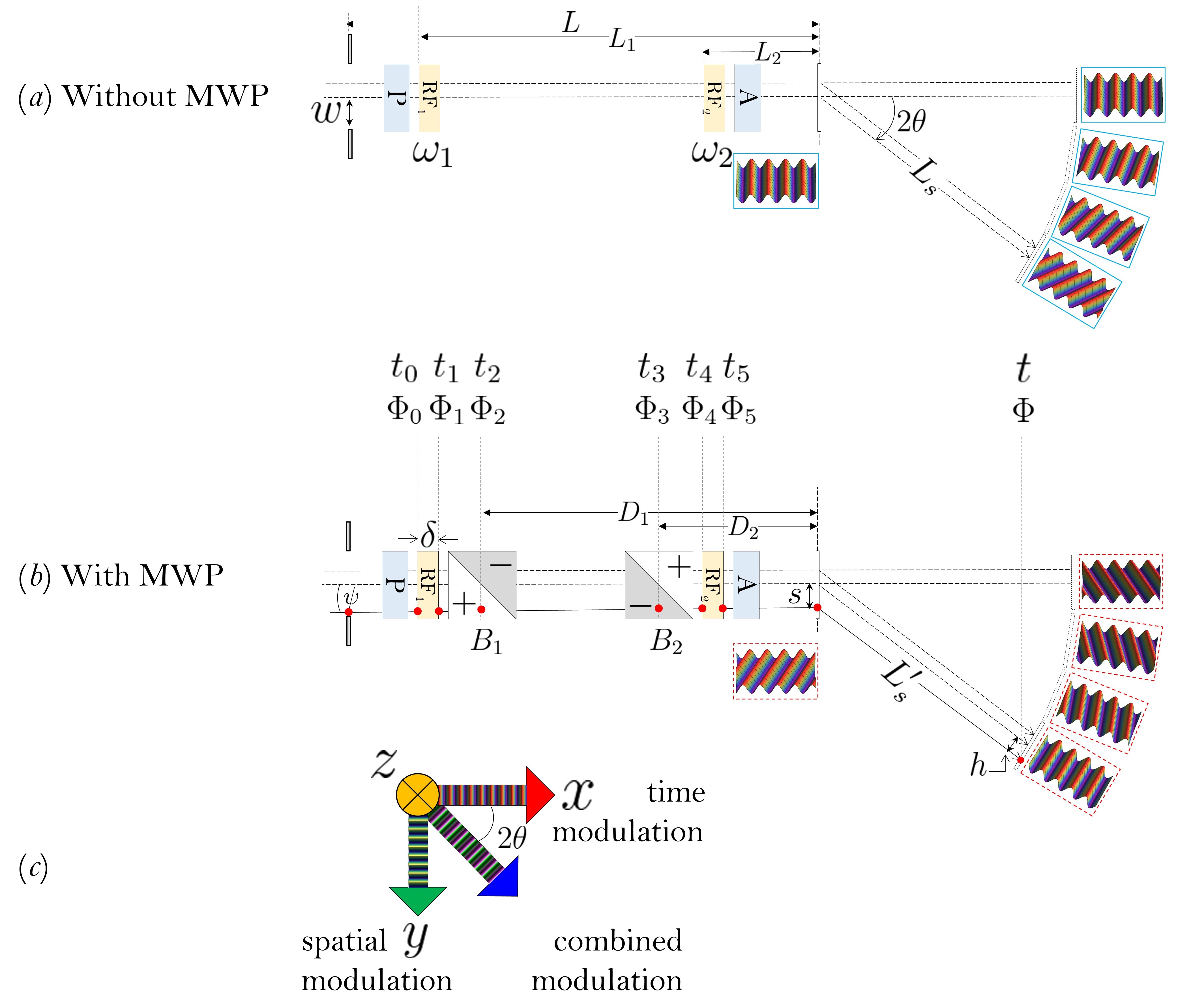}
	\caption { (a) The original MIEZE setup.  (b) The MIEZE setup combined with magnetic Wollaston prisms (MWPs) ro correct the linear Larmor phase from the sample. (c) The schematic of the combination of intensity modulations in time and space domain. The light yellow boxes are the RF neutron spin flippers with angular frequency of $\omega_{1,2}=2\pi f_{1,2}$.  The MWPs are denoted by the light gray boxes with a field of $B_{1,2}$ respectively. $L_{1,2}$ and $D_{1,2}$ are the distances from the RF flippers and MWPs to the sample. $P$ and $A$ denote the neutron spin polarizer and analyzer. The neutron intensity modulations are developed from the RF flippers in the time domain and a snapshot of the intensity modulation at a specific time has been shown at the sample and detector position, for the case of without (a) and with (b) the MWPs. $s, h$ are the position where the neutrons are scattered and captured in the sample and detector respectively. $w$ is the transverse offset of the neutron's trajectory at the entrance of the slit located at $x=L$, which is used to define the divergence of the incoming beam. $2W, 2S$ and $2H$ are the full size of the slit, sample and detector respectively. $\Phi_i$ is the phase of neutron spin at various positions and time ($t_i$). For demonstration purposes, the dashed trajectories are the ideal case where both the incoming and outgoing neutron beam are parallel to the center trajectory. The solid trajectory shows the actual case in the experiment, which is used in the  following calculations. The red dots along the neutron trajectory are the intersection points with variosu components of the setup. A flat and purely elastic sample with no thickness is assumed for all the calculations. }
	\label{setup}
\end{figure*}

An alternative approach to allow for complicated sample environment and depolarizing sample is to use Modulation of Intensity Emerging from Zero Effort (MIEZE) \cite{RN2366}, which has been demonstrated or routinely operated at the instruments RESEDA of FRM-II\cite{RN2372},VIN ROSE of J-PARC \cite{HINO2013136}, LARMOR of ISIS \cite{larmor}, and ORNL \cite{BRANDL20121, dadisman, zhao}.  It requires two neutron resonance radio-frequency (RF) spin flippers before the sample operating at different RF frequencies. With such spin flippers, the polarization vector of the neutron beam can be modulated, with their phase determined by the phase angle of the RF signal \cite{RN2366}. After passing through the neutron polarization analyzer, which picks up the polarization vector along one direction, the polarization modulation can be converted into intensity modulation before interacting with the sample, as shown in Figure \ref{setup} (a). Because the ensemble of neutrons doesn't carry any net polarization anymore, the neutron beam is no longer sensitive to the factors that can depolarize the neutron in the sample or sample environment, which allows MIEZE for complicated sample environment. When neutrons interact with the sample, their energy will be changed, which may cause a time advance or delay of their arrival at the detector. Therefore, the contrast of the modulation will be changed, from which the intermediate scattering function of the sample can be deduced. 

However, one of the problems of MIEZE is the Larmor phase aberration which occurs due to the finite transverse size of the sample.  This problem originates from the variation of neutron time-of-flight from the sample to the detector. To make it clearer, as shown in  Figure \ref{setup} (a), and for demonstration purpose, both the incoming and scattered neutrons are parallel to the beam center line with no wavelength dispersion. The wavefront of the intensity modulation is perpendicular to the beam direction at the sample position, as well as the detector position in the transmission geometry. But for the detector at an angle ($2\theta$), the arrival time of a neutron will vary depending on the position of the scattering event at the sample position ($s$). The consequent wavefront of the scattered neutrons will be twisted and misaligned with the detector plane, which will introduce phase aberration when binning the detector data spatially to improve statistics. Also, the phase aberration will increase as scattering angle, which means the resolution of MIEZE especially at large scattering angles will be greatly limited. Without resolving this problem, it is challenging to further increase the energy resolution especially at large scattering angles without compromising the size of the sample and thus neutron intensity. It has been suggested by Martin \cite{RN2367} that the detector can be simply configured to be perpendicular to the incident beam direction to avoid such phase aberration. However, this approach provides a complete correction only in the ideal case when both the incident and scattered neutrons are parallel beams, which follow the center line shown in Figure \ref{setup} (a). In practice, the incident neutrons come with divergence and the scattering from the sample also creates divergence, which means this phase aberration cannot be corrected by simply tilting the detector or post data reduction. We have also shown in ref. \cite{tilt} that the optimum detector setting needs to be perpendicular to the trajectory of the scattered neutrons. As stated by Brandl \cite{RN2371} and Weber\cite{RN2370}, tilting the sample with respect to the beam can also change the resolution function of the MIEZE because it changes the neutron time-of-flight from the sample to the detector. However, tilting a flat sample is only practical within limits and it will significantly limit the choices of sample. Though it has been shown the resolution function of MIEZE can also be manipulated by physically tilting the RF flipper \cite{tilt}, its performances are sensitive to the beam divergence.

In this paper, I propose using two magnetic Wollaston prisms (MWPs) to tune the resolution function of MIEZE such that the sensitivity of the setup can be maximized at any scattering angle of interest without any physical tilting of the sample or the RF flipper. MIEZE, with two RF flippers operated at different frequencies $\omega_{1,2}$, generates a Larmor phase $\Phi(t)$ with a linear gradient in the time domain along $x$ direction yielding a modulation, as shown in Figure \ref{setup} (c). The Larmor phase contributed by the RF flippers for each neutron captured at the detector position is $\Phi(t)=2\Delta \omega t=2(\omega_2-\omega_1)t$, where $t$ is the clock of the detector. The MWPs energized with opposite magnetic fields can generate a linear Larmor phase gradient and hence modulation along the transverse direction ($y$), as shown in Figure \ref{setup} (c). With the addition of the MWPs, the gradient direction of the total Larmor phase can be changed from $x$ direction to any direction in the $xy$ plane by picking the right combination of the phase gradients in the two directions, as shown in Figure \ref{setup}(b) and (c). Since the wavefront of the intensity modulation is given by the direction of the phase gradient, the principle of this method is to tune the direction of the phase gradient such that the wavefront is parallel to the detector plane, where phase aberrations can thus be minimized. 

\section{Larmor Phase of MIEZE with Magnetic Wollaston Prisms}
Before showing the total Larmor phase a neutron picks up through the setup in Figure \ref{setup} (b), the Larmor phase contributed by either a RF flipper or a MWP will be briefly discussed. As given by R. Golub \textit{et. al.} \cite{RN2373}, the Larmor phase generated by a neutron RF flipper with an angular frequency of $\omega$ is given by $\Phi_f=2\Phi_{rf}(t_i)+\omega t_{\pi}-\phi_i$. $\Phi_{rf}=\omega t_i$ is the phase of the RF flipper upon the entrance of the neutron at $t_i$, $t_{\pi}$ is the time neutron stays inside the flipper $t_{\pi}=\frac{\delta}{v \cos\psi}$ with $\psi$ being the divergence angle, $\delta$ being the thickness of the flipper and $v$ being the neutron velocity, and $\phi_i$ is the phase neutron carries before entering the RF flipper. For a MWP, as given by F. Li \cite{RN2374,RN2375,RN2376}, the Larmor phase generated ($\Phi_m$) depends on the magnetic field integral along the neutron's trajectory $\Phi_m(B,y,\psi)=\frac{2\gamma }{v}By[\cot\beta+(\cot\beta)^2\psi]$, where $\gamma$ is the gyromagnetic ratio of neutron, $B$ is the magnetic field of the MWP, $\beta$ is the inclination angle of the hypotenuse of the MWP with respect to the beam and $y$ is the transverse coordinate of the intersection point of the beam with the midplane of the MWP, as denoted by the red dotes in Figure \ref{setup} (b). With $\Phi_m$, the MWP can be taken as a device with no thickness along the beam. With each neutron's transverse coordinates, divergence and velocity known, the Larmor phase picked up inside the MWP can be calculated. While the inclination angle of the MWP ($\beta$) can be any value, in the following calculations, it will be assumed to be $45^{\circ}$. For a neutron trajectory incident upon the sample with an offset of $s$ and divergence of $\psi$ as indicated by the solid black line in Figure \ref{setup}(b), the intersection points with each device are denoted by the red dots. The coordinates of the intersection in time and space and the corresponding Larmor phase have been calculated and shown in Table \ref{tab}, with which the Larmor phase can be calculated and propagated to the detector, yielding,

\begin{strip}
\begin{equation} 
\normalsize
\begin{split}
\Phi(t)=&\underbracket[0.5pt][4pt]{2(\omega_2-\omega_1)t}_{\circled{1}}+\frac{1}{v}\Bigg[\underbracket[0.5pt][4pt]{2(\omega_1L_1-\omega_2L_2)\sqrt{1+\psi^2}}_{\circled{2}}+(\omega_1-\omega_2)\bigg(-\delta\sec\psi+2\sqrt{\underbracket[0.5pt][4pt]{L_s^2}_{\circled{2}}+\underbracket[0.5pt][4pt]{h^2+s^2-2s(h\cos2\theta+L_s\sin2\theta)}_{\circled{3}}}\bigg)\\
&+2\gamma\bigg(\underbracket[0.5pt][4pt]{(B_1-B_2)s}_{\circled{4}}+\underbracket[0.5pt][4pt]{(B_1D_1-B_2D_2)\psi}_{\circled{5}}\bigg)(1+\psi)\Bigg]\\
%\approx&\underbracket[0.5pt][4pt]{2(\omega_2-\omega_1)t}_{\text{\normalsize Time modulation}}+\frac{2}{v}\bigg(\underbracket[0.5pt][4pt]{ \omega_1(L_1+L_s)- \omega_2 (L_2+L_s)}_{\text{\normalsize Time focusing}}+\underbracket[0.5pt][4pt]{\gamma\psi(B_1D_1-B_2D_2) }_{\text{\normalsize Space focusing}}+ \underbracket[0.5pt][4pt]{s \Big(\gamma  (B_1-B_2)- (\omega_1-\omega _2)\sin 2 \theta \Big)}_{\text{\normalsize Focusing condition}}\bigg)
%
\approx&2(\omega_2-\omega_1)t+\frac{2}{v}\Bigg[\underline{ \omega_1(L_1+L_s)- \omega_2 (L_2+L_s)}+\underline{\gamma\psi(B_1D_1-B_2D_2) }+ s\bigg (\underline{\gamma  (B_1-B_2)- (\omega_1-\omega _2)\sin 2 \theta}\bigg )\Bigg]
\end{split}
\label{phase}
\end{equation}
\end{strip}
\begin{table}
\center
\caption{The propagation of the Larmor phase in the proposed MIEZE setup. Please refer to Figure \ref{setup} (b) for the detailed definitions of the parameters.}
%\footnotesize

\begin{tabular}{l | c | l}
\hline
\hline
Position              &  Time ($t_i$)  &  Neutron Phase ($\Phi_i$) \\
\hline

Entrance of RF$_1$ & $t_0$  & $\Phi_0=0$   \\
Exit of RF$_1$ & $t_1$ &  $\Phi_1=2\omega_1(t_0+\frac{\delta}{2v\cos{\psi}})-\Phi_0$   \\

Center of MWP$_1$ & $t_2$ & $\Phi_2=\Phi_m(B_1,y_1,\psi)+\Phi_1$   \\
Center of MWP$_2$ & $t_3$  & $\Phi_3=\Phi_m(B_2,y_2,\psi)+\Phi_2$   \\

Entrance of RF$_2$ & $t_4$  & $\Phi_4=\Phi_3$   \\
Exit of RF$_2$ & $t_5$  & $\Phi_5=2\omega_2(t_4+\frac{\delta}{2v\cos{\psi}})-\Phi_4$   \\
Detector& $t$  & $\Phi=\Phi_5$   \\  
\hline
\hline
\end{tabular}
\label{tab}
\begin{tabular}{l}
$t_0=t-\frac{1}{v}\bigg(L_1\sqrt{1+\psi^2}+L'_s\bigg)$\\
$y_1=-D_1 \psi-s $ \\
$y_2=-D_2\psi-s$ \\
$t_4=t-\frac{1}{v}\bigg(L_2\sqrt{1+\psi^2}+L'_s\bigg)$\\
$L'_s=\sqrt{h^2+L_s^2+s^2-2s(h\cos{2\theta}+L_s\sin{2\theta})}$\\
\end{tabular}
\end{table}

In the calculation, the divergence angle can be explicitly defined as $\psi=\frac{w-s}{L}$, where $L$ is the distance of the slit to the sample position and $s$ and $w$ are the position where the scattering takes place in the sample and the displacement of the neutron trajectory at the slit position respectively. The terms labeled with different numbers in Equation \ref{phase} represent the major terms that contribute to the Larmor phase. Whereas the terms not labeled are higher order terms and thus are negligible. The terms  labeled in \circled{1} denote the intensity modulation produced by the RF flippers. The Larmor phase aberration, due to the dispersion of neutron velocity ($v$), transverse offset of scattering ($s$) and the distribution of scattering angles ($2\theta$), is explicitly given as \circled{2} and \circled{3}. The Larmor phase generated by the MWP is provided by \circled{4} and \circled{5}, which are time independent. Term \circled{1} is the main MIEZE signal of interest, term \circled{2} and \circled{3} are the aberration terms contributed from the sample size, scattering angle, neutron velocity etc, which needs to be minimized by the correction. In a MIEZE setup, the sample size ($\sim2cm$) and the detector size  ($\sim10cm$) are relatively smaller than the length of the scattering arm  ($L_s>2m$), and also the divergence of the beam is small. Therefore, Equation \ref{phase} can be further expanded in $s, h$ and $\psi$. With the lowest orders summerized in Equation \ref{phase}, for the velocity dependent terms, we will minimize their contributions by setting each of the underlined combinations to zero, yielding,

\begin{align}
\text{Time focusing: } &\omega_1(L_1+L_s)=\omega_2(L_2+L_s) \label{mieze}\\
\text{Space focusing: }&B_1D_1=B_2D_2 \label{semsans} \\
\text{Hybrid focusing: }&B_1=B_2+\frac{1}{\gamma}(\omega_1-\omega_2)\sin2\theta \label{mwp}  
\end{align} 
Equation \ref{mieze} is termed the time focusing condition or the MIEZE condition. In a conventional MIEZE setup without MWPs, it can remove the velocity dependence of the Larmor phase for the directly transmitted neutron beam. Equation \ref{semsans} can ensure the Larmor phase the neutron spin accumulates through the MWPs is independent of the beam divergence. With such condition, the wavefront of neutron spin can be tilted with respect to the incident beam direction by the same amount regardless of their divergence angle ($\psi$). For a given scattering angle (2$\theta$) and RF frequency ($\omega_{1,2}$), to ensure the wavefront can be tilted by the right amount , Equation \ref{mwp} needs to be satisfied. Since Equation \ref{mwp} involves the intensity oscillations in both time and space domain, I will refer it as the hybrid focusing condition in the following discussion. Equation \ref{semsans} and \ref{mwp} together yield the optimum field inside required for the MWPs to compensate the linear phase aberration due to the transverse dimension of the sample in a regular MIEZE. In combination with Equation \ref{mieze}, the Larmor phase of neutron spin at the detector position would be independent of the neutron velocity and beam divergence to the first order, which is the prerequisite condition to maiximize the contrast of the intensity modulation.

To understand the principle of this method more intuitively, the introduction of the two MWPs can generate a field integral gradient along the transverse direction ($y$) as shown in Figure \ref{setup} (c), which is perpendicular to the phase gradient direction generated by the RF flippers ($x$). With the two phase gradients, the combined gradient direction of the Larmor phase can be manipulated to any direction of interest. Since the gradient direction is always perpendicular to the wavefront of the intensity modulation, it can be tuned such that it is perpendicular to the surface of the detector. In this case, the wavefront will be parallel to the detector and the phase aberrations can be minimized. We can also see that, when the scattering angle is small ($2\theta\approx0$), no field is required for the MWPs ($B_{1,2}=0$). As also shown by Equation \ref{mwp}, the field required for the MWPs is independent of neutron wavelength, which makes it suitable for pulsed neutron beam. 

\section{The Phase Correction Calculations of MIEZE}

With the Larmor phase known as Equation \ref{phase}, the time modulations measured by the detector can be calculated by integrating over the beam size ($w$) and thus divergence ($w=L\psi+s$) and sample size ($s$) in Figure \ref{setup}, as given in Equation \ref{fringe}. Ultimately, the detector will measure a 3-dimentional data set with one dimension denotes the time and the other two denote the pixel coordinates of the detector. Since MIEZE is a low signal measurement, the data is then integrated over the detector spatially to improve the statistics, after which the data becomes a 1-dimensional (1D) intensity oscillation in the time domain. The detector thickness is assumed to be zero in the calculation. The contrast of the modulation can therefore be given by the amplitude of the 1D oscillation.  A higher contrast will yield a higher sensitivity to a small change in neutron energy. Therefore, the dependence of the contrast as a function of the scattering angle denotes the resolution function of the MIEZE setup.
\begin{equation} 
\small
R(2\theta,t)=\int_{H}^{H}\int_{-S}^{S}\int_{-W}^{W} \cos\Big(\Phi(2\theta,\omega_1,B_1,w,s,h,t) \Big)\,dw{\;}ds{\;}dh
\label{fringe}
\end{equation}

\subsection{The Phase Aberrations of a Regular MIEZE}
\begin{figure}[ht]
	\centering
	\includegraphics[width=1\linewidth, clip]{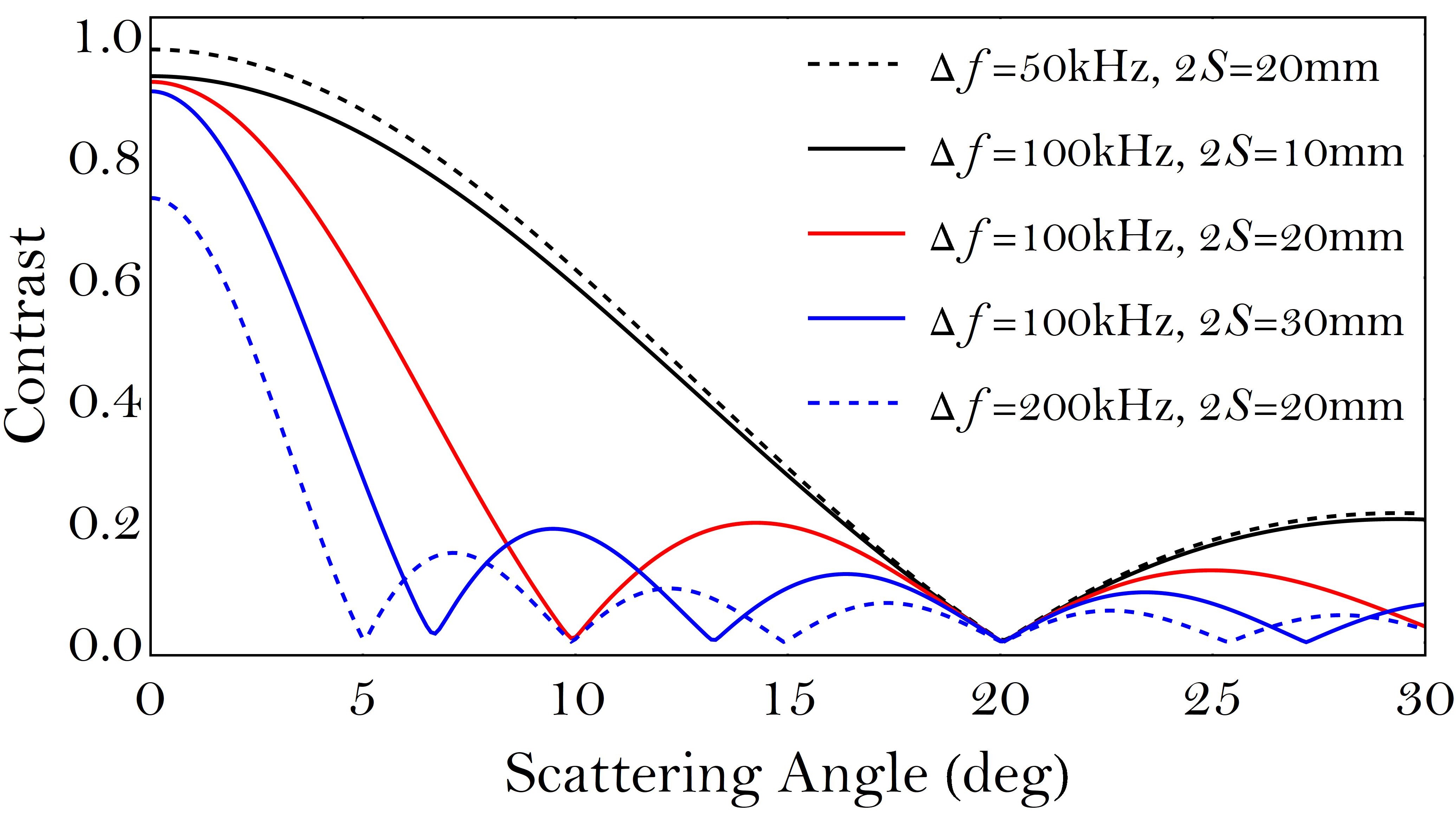}
	\caption { The contrast of the modulation as a function of the scattering angle $2\theta$ when changing the frequency difference between the two RF flippers ($\Delta f$) and the size of the sample (2$S$). The two RF flippers are maintained in time focusing.}
	\label{aberration}
\end{figure}
First, to demonstrate the phase aberration problem of a regular MIEZE, calculations have been performed by setting the field in the MWPs to zero in Equation \ref{phase}. In the calculation, the parameters are picked such that $L_1=4m, L_2=2m, L_s=2m, 2H=100mm, \delta=30mm$, which are very close to a realistic MIEZE instrument. For demonstration purposes, a parallel beam is used for this calculation. The contrast of the intensity modulation has been plotted in Figure \ref{aberration} as a function of scattering angle ($2\theta$) for various frequencies ($\Delta f$) and sample size ($2S$). As we can see, the peak contrast of the resolution function is always centered at zero scattering angle and the contrast of the modulation decreases rapidly as scattering angle increases especially for large sample size (2$S$) and high RF frequency differences ($\Delta f$). For example, for $\Delta f$=200kHz and $2S$=20$mm$, it is almost impossible to measure the scattering at a scattering angle of 5$^{\circ}$ with a vanishing contrast. This can also be intuitively understood by the illustration in Figure \ref{setup} (a) that the  neutrons scattered at different locations of the sample cannot arrive at the detector plane at the same time such that aberrations will occur when integrating over the detector.
\subsection{The Scattering Angle Dependence of the Phase Correction}
\begin{figure}[ht]
	\centering
	\includegraphics[width=\linewidth, clip]{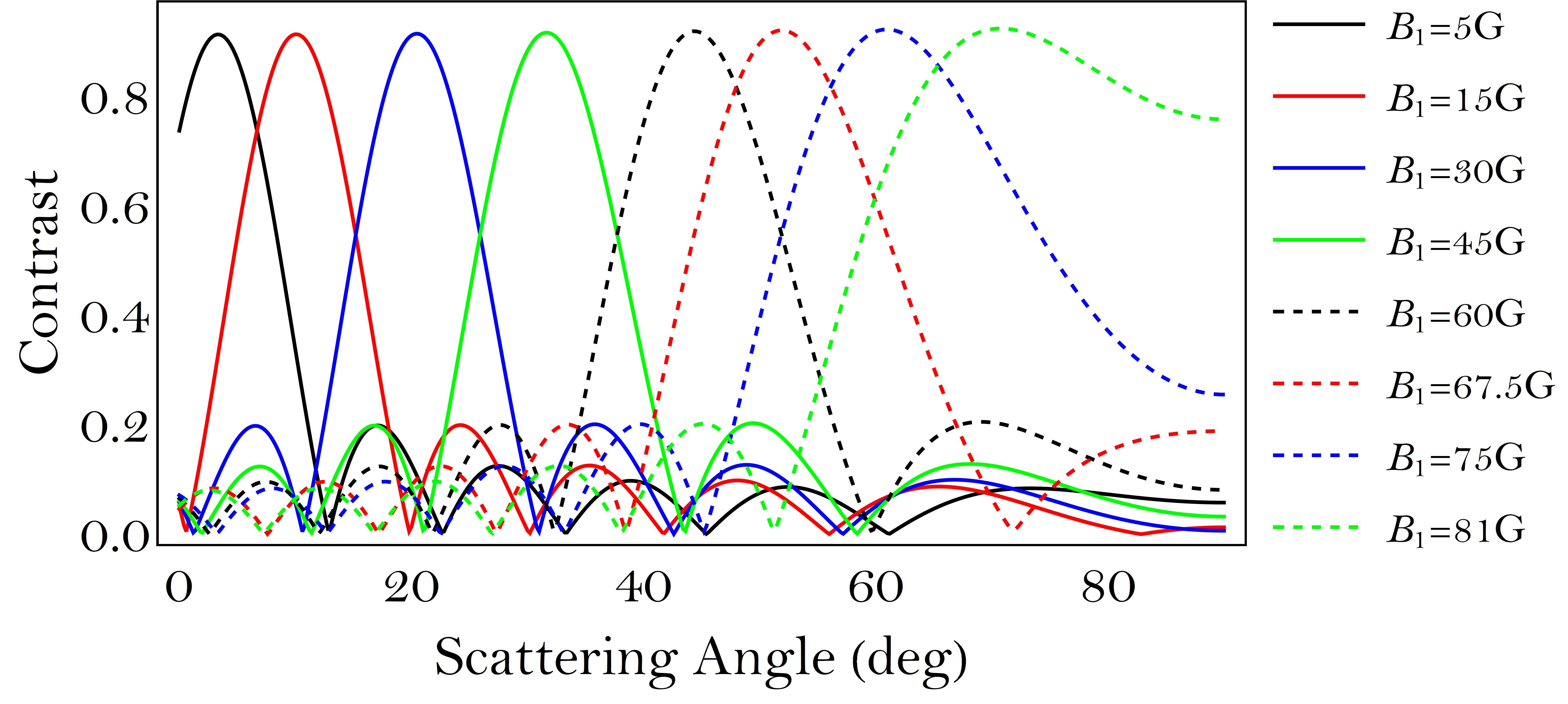}
	\caption {The resolution function of MIEZE as a function of the scattering angle with the MWPs energized according to Equation \ref{mwp}. Different lines represent the resolution function of various fields inside the MWP. In this calculation, $\lambda=5.5\AA,\Delta  f=100$ kHz,  $ 2S=20mm, L=5m, 2W=40mm$ and $2H=100mm$.}
	\label{angle}
\end{figure}

The resolution function of the setup has been calculated and plotted in Figure \ref{angle}, assuming that the MWPs are energized according to Equation (\ref{semsans}, \ref{mwp}). For example, to recover the contrast of the modulation at a scattering angle of 20$^{\circ}$ at a RF frequency difference of $\Delta f=100kHz$, the field required in the first MWP is ~30G. As one can see, with the correction provided by the MWPs, the resolution function can be tuned, and the peak contrast can be shifted to any scattering angle. Namely, the resolution function of the MIEZE setup can be optimized by the MWPs such that the highest sensitivity can be fully recovered at any scattering angle of interest. This would allow the MIEZE instrument to reach any scattering angle without compromising the contrast of the intensity modulation or the size of the sample.

\subsection{Frequency Dependence of the Phase Correction}

\begin{figure}[ht]
	\centering
	\includegraphics[width=\linewidth, clip]{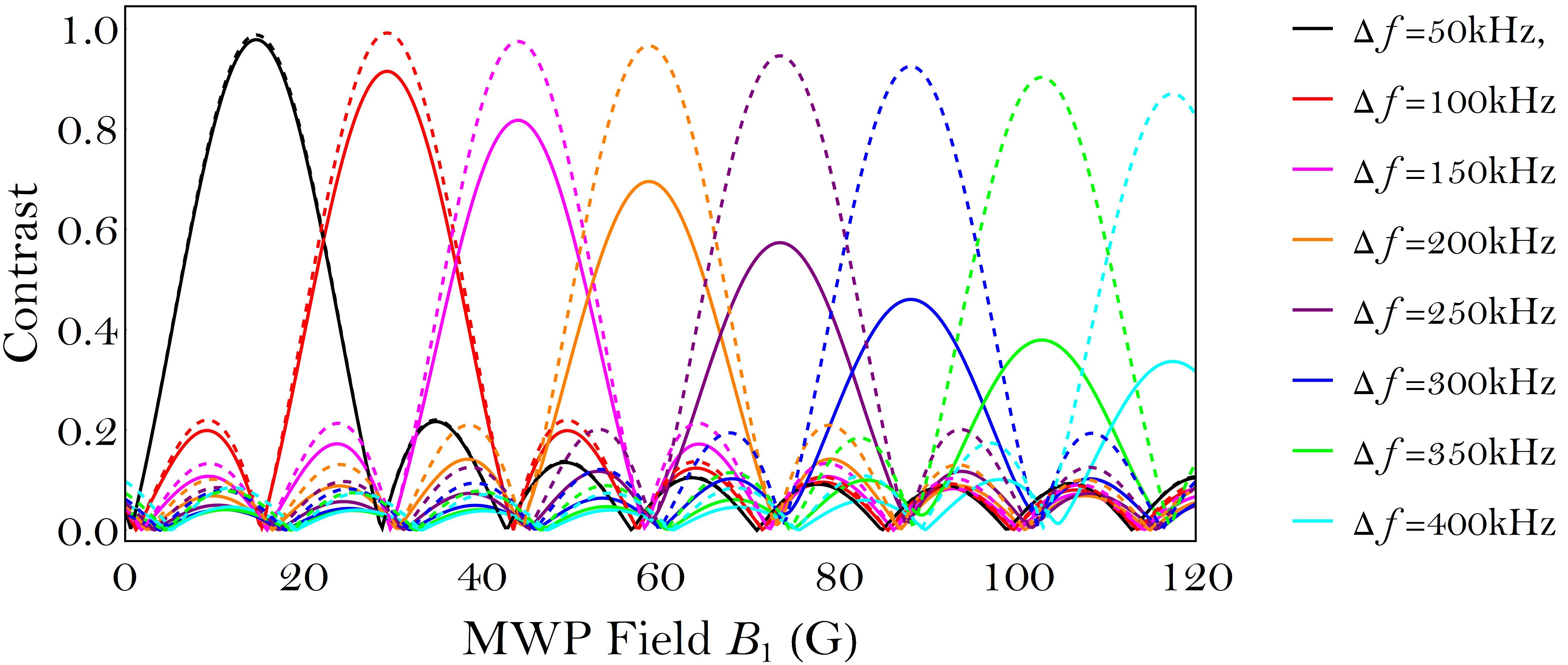}
	\caption {The variation of the  resolution function for a series of RF frequencies at a fixed scattering angle of $20^{\circ}$.  Solid and dashed lines represent a detector size of $2H=100mm$ and 50$mm$ respectively.}
	\label{fre}
\end{figure}

At a fixed scattering angle, the frequency dependence of the resolution function has been calculated and shown in Figure \ref{fre}. In this plot, the scattering angle is fixed to be $2\theta=20^{\circ}$ and the magnetic fields inside the MWPs are scanned such that the contrast of the intensity modulation can be optimized for each frequency setting. As expected from Equation \ref{mwp}, as the frequency increases, the field required for the MWPs increases. However, as shown in Figure \ref{fre}, the peak value of the resolution function will decrease as the increase of the RF frequency. This is because, as the modulation frequency increases, the period of modulations in the temporal and spatial space combined gets smaller as a general trend. With the transverse size of the detector being fixed, it will integrate over a larger phase volume of the modulation. Therefore, more aberrations will be introduced causing a lower contrast and thus sensitivity. When halving the size of the detector from 100$mm$ to 50$mm$, as shown in Figure \ref{fre}, the peak value of the resolution function will be recovered. So, when optimizing a MIEZE instrument, the phase volume that the detector occupies in the time-space domain needs to be evaluated.

\subsection {Wavelength Dependence of the Phase Correction}
\begin{figure}[h]
	\centering
	\includegraphics[width=0.7\linewidth, clip]{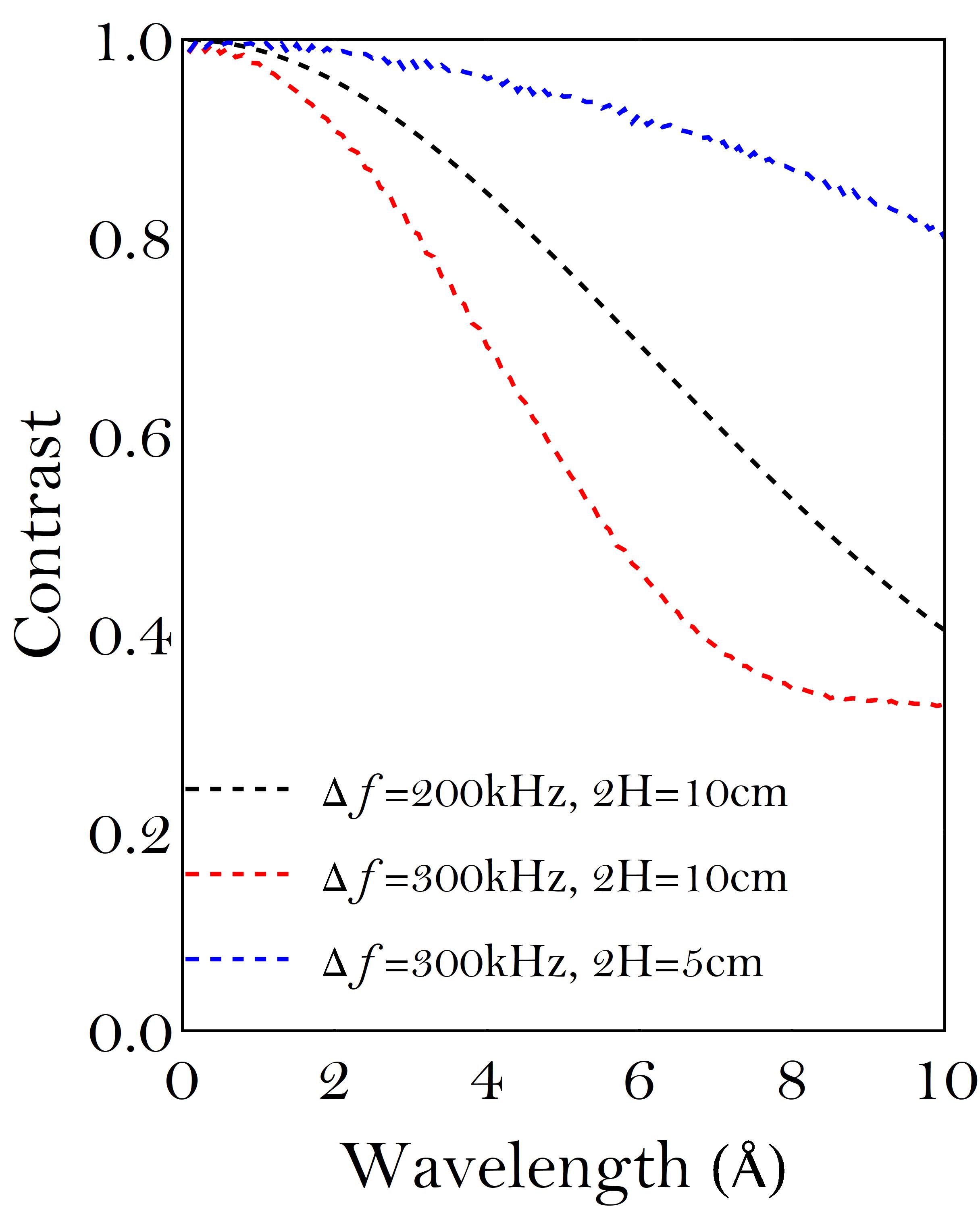}
	\caption {The variation of the peak value of the resolution function as a function of wavelength for various frequencies and detector size. For each case, the field in the MWP has been optimized according to Equation \ref{mwp}. The parameters for the calculation are $2\theta=20^{\circ}, 2W=40mm$ and $2S=20mm$. }
	\label{tof}
\end{figure}

It has been shown by Equation \ref{mwp} that the optimum fields of the MWPs are wavelength independent. The applicability of this method to a pulsed neutron source is now being calculated. The calculations are performed with two RF frequencies ($\Delta f=200, 300$kHz) and two different detector size ($2H=100mm, 50mm$). The scattering angles is fixed to be $2\theta=20^{\circ}$. For each setting, the fields of the MWPs are optimized and the neutron wavelength is swept to obtain the resolution function. The results are shown in Figure \ref{tof}. As one can see, the contrast of the modulation will drop towards longer wavelength and higher frequency. As discussed in the previous section, as the RF frequency and the neutron wavelength increase, the volume size of the phase space that the detector measures will be increased too, leading to a phase aberration. It can be compensated by reducing the detector size, as shown in Figure \ref{tof}. 

\section {Discussions and Conclusions}

Using two magnetic Wollaston prisms, I have presented an approach to correct the first order Larmor phase aberration of MIEZE caused by the transverse size of the sample. With this approach, the resolution function of MIEZE can be modified such that the contrast of the intensity modulation can be maximized at any scattering angle of interest regardless the beam divergence and neutron wavelength. But this approach still limits MIEZE to be used at a small solid angle. I have shown that a large detector would take a large phase volume, which will potentially cause more phase aberration. But a large pixelated detector is still of great value especially if the phase variation between pixels on the detector surface is known. With such infomation, the phase aberration can be further reduced by post data treatment procedure.

\section{Acknowledgments}
This work was sponsored by the Laboratory Directed Research and Development Program of Oak Ridge National Laboratory, managed by UT-Battelle, LLC, for the U. S. Department of Energy. This material is based upon work supported by the U.S. Department of Energy, Office of Science, Office of Basic Energy Sciences under contract number DE-AC05-00OR22725. I am also grateful to Steven Parnell (Delft University of Technology), Roger Pynn, Steve Kuhn (Indiana University Bloomington) and Georg Ehlers (Oak Ridge National Laboratory) for their help with the manuscript.

\bibliographystyle{elsarticle-num}

\bibliography{ref}

\end{document}